\documentstyle[11pt,psfig]{article}
\def\reference{\parskip 0pt\par\noindent\hangindent 0.5 truecm}

\def\Mpc{\hbox{$\rm\thinspace Mpc$}}

\def\km{{\rm\thinspace km}}
\def\s{{\rm\thinspace s}}

\def\hMpc{\hbox{$\thinspace h^{-1}\Mpc$}}
\def\kms{\hbox{$\km\s^{-1}\,$}}
\def\kmsMpc{\hbox{$\kms\Mpc^{-1}$}}

\begin{document}
%
%
\title{The Large Scale Distribution of Galaxies in the Shapley Supercluster}
%

\author{Michael J. Drinkwater$^{1}$, 
 Quentin A. Parker$^{2}$, 
 Dominique Proust$^{3}$,\and
 Eric Slezak$^{4}$,  Hern\'an Quintana$^{5}$
} 

\date{Submitted to PASA 2003 Sep 18}
\maketitle

{\center
$^1$ Department of Physics, University of Queensland, QLD 4072, Australia,
\\mjd@physics.uq.edu.au\\[3mm]
$^2$ Department of Physics, Macquarie University, NSW 2109, Australia,
and Anglo-Australian Observatory, PO Box 296, Epping NSW 1710,  Australia,
\\qap@ics.mq.edu.au\\[3mm]
$^3$ Observatoire de Paris-Meudon, F92195 Meudon CEDEX, France
\\Dominique.Proust@obspm.fr\\[3mm]
$^4$ Observatoire de Nice, 06304 Nice CEDEX4, France, 
Slezak@obs-nice.fr\\[3mm]
$^5$ Departamento de Astronomia y Astrofisica, Pontificia Universidad
Cat\'olica de Chile, Casilla 104, Santiago 22, Chile,
hquintan@astro.puc.cl\\
}

\begin{abstract}
We present new results of our wide-field redshift survey of galaxies
in a 182 square degree region of the Shapley Supercluster (SSC) based
on observations with the FLAIR-II spectrograph on the UK Schmidt
Telescope (UKST). In this paper we present new measurements to give a
total sample of redshifts for 710 bright ($R\leq 16.6 $) galaxies, of
which 464 are members of the SSC ($8000<v<18000\kms $). Our data
reveal that the main plane of the SSC ($v\approx 14500\kms $) extends
further than previously realised, filling the whole extent of our
survey region of 10 degrees by 20 degrees on the sky (35 Mpc by 70
Mpc, $H_0=75\kmsMpc$). There is also a significant structure
associated with the slightly nearer Abell 3571 cluster complex
($v\approx12000\kms$) with a caustic structure evident out to a radius
of 6 Mpc.  These galaxies seem to link two previously identified
sheets of galaxies and establish a connection with a third one at
$\overline V$= 15000\kms near $R.A.= 13h$. They also tend to fill the
gap of galaxies between the foreground Hydra-Centaurus region and the
more distant SSC. We calculate galaxy overdensities of $5.0\pm0.1$
over the 182 square degree region surveyed and $3.3\pm0.1$ in a 159
square degree region excluding rich clusters. Over the large region of
our survey the inter-cluster galaxies make up 46 per cent of all
galaxies in the SSC region and may contribute a similar amount of mass
to the cluster galaxies.
\end{abstract}

\bigskip

%
%
\newpage
\section{Introduction}

In the past few decades, large galaxy redshift surveys have revealed
structures on ever-increasing scales. The largest structures found are
superclusters, collections of thousands of galaxies with sizes as
large as 100 Mpc. The mere existence of these structures places
important constraints on theories of the formation of galaxies and
clusters. The Shapley supercluster, the subject of this paper,
is one of the most massive concentrations of galaxies in the local
universe (Scaramella et al.\ 1989; Raychaudhury 1989), so it is also
of particular interest to consider its effect on the dynamics of the
Local Group.

The Shapley supercluster (SSC) is a remarkably rich concentration of
galaxies centred around R.A.$=13^h25^m$ Dec. $= -30^{\circ}$ which has
been investigated by numerous authors since its discovery in 1930 (see
Quintana et al. 1995). It consists of many clusters and groups of
galaxies in the redshift range $0.04<z<0.055$. The SSC lies in the
general direction of the dipole anisotropy of the Cosmic Microwave
Background (CMB), (Smoot et al. 1992), and is located 130\hMpc\ beyond
the Hydra-Centaurus supercluster (itself $\simeq 50$\hMpc\ away from
us).  Quintana et al.\ (1995) estimated that for $\Omega_{o}= 0.3$ and
$H_{o}$= 75\kmsMpc the gravitational pull of the supercluster may
account for up to 25\% of the peculiar velocity of the Local Group
required to explain the CMB dipole anisotropy in which case the mass
of the supercluster would be dominated by inter-cluster dark matter.
A major study of the SSC was made by Bardelli et al.\ (2000, 2001 and
references therein) who also studied inter-cluster galaxies in the
core region of the supercluster. They derived a total mean overdensity
of the SSC of $N/\overline N \sim 11.3$ on a scale of $10.1h^{-1}$ Mpc
and found that the central part of the supercluster contributes about
26\kms\ to the peculiar velocity of the Local Group.


The early studies of the Shapley supercluster were limited
(primarily by observational constraints) to the various rich Abell
galaxy clusters in the region, but this might give a very biased view
of the overall supercluster as they represent only the most
concentrated peaks in the luminous matter distribution. We have
commenced an investigation into the larger scale distributions of
galaxies throughout the entire SSC region and close environs using
data from wide-field multi-fibre spectrographs such as the FLAIR-II
system on the U.K. Schmidt Telescope (UKST) of the Anglo-Australian
Observatory (Parker \& Watson, 1995, Parker 1997). With such
multiplexed facilities we are able to measure many more galaxy
redshifts over large angular extents and obtain a more complete
picture of the composition and disposition of galaxies in the entire
supercluster. 

A preliminary analysis based on 301 new, bright ($R\leq 16$) galaxy
redshifts obtained with FLAIR-II in a 77~deg$^{2}$ region of the
supercluster has already been published by our group (Drinkwater et
al.\ 1999, hereafter D99). The measured galaxies were uniformly
distributed over the selected inter-cluster regions, and most ($230
\equiv 75\%$) were found to be members of the supercluster. The D99
sample traced out two previously unknown sheets of galaxies which
appeared to link various Abell clusters in the supercluster. We also
found that in a 44 deg$^{2}$ sub-area of the supercluster {\em
excluding} the Abell clusters, these sheets alone represent an
overdensity of a factor of $2.0\pm0.2$ compared to a uniform galaxy
distribution.  Within the initial survey area of D99 the new galaxies
contribute an additional 50\% to the known contents of the Shapley
supercluster in that region, implying a corresponding increase in
contribution to the motion of the Local Group. A much larger galaxy
sample over a similar area of sky was presented by Quintana et al.\
(2000). They measured nearly 3000 galaxy redshifts in an area of about
$12\deg \times 6\deg$. They estimated the upper bound on the mass of
the central region and found the overdensity to be substantial, but
still only able to contribute to a small fraction
($3\Omega_m^{-0.4}$\%) of the observed motion of the Local Group
(Reisenegger et al.\ 2000).

In this paper we present radial velocities for an additional 409
bright ($R\leq16.5$) galaxies spread over an extended region to the
East, West and South of the main SSC concentration. We analyse a
combined magnitude-limited sample from this paper and D99 of 710
galaxies with measured redshifts, of which 464 (65 per cent) are
members of the supercluster based on their redshift range ($8,000 <
cz < 18,000\kms$). They seem to link the two previously identified
galaxy sheets found in D99 and also establish a connection with a
third feature at $\overline V$= 15000\kms near R.A.=~13h. The new
sample also tends to fill a previous gap in the galaxy distribution
between the foreground Hydra-Centaurus region and the SSC. Future
study of the SSC will cover even larger regions using data from the
FLASH survey of Kaldare et al.\ (2003) (also with the UKST/FLAIR-II
system) and the 6dF survey of the southern sky currently in progress
(Watson et al.\ 2000).

The observed galaxy sample and observations are described in
Section~2, the results are presented in Section~3 and a brief
discussion of the significance of these new measurements is given in
Section~4.  A full and detailed analysis and interpretation of the new
redshifts from D99, this work and a compilation of all literature
redshifts in the SSC is the subject of a separate paper (Proust et
al., in preparation). Unless otherwise noted in this paper we use a
Hubble constant of $H_0=75\kmsMpc$ giving a distance of 200 Mpc to the
main supercluster ($cz=15,000\kms$) so 1 degree subtends a distance of
3.5 Mpc.

\section{Observations and Data Reductions}

In 1997 we began a concerted campaign to obtain complete samples of
galaxies down to the same magnitude below $L_*$ for constituent
clusters and inter-cluster regions of the SSC using the FLAIR-II
system at the UKST. During the period 1997-2000 we subsequently
observed more than 700 selected galaxies over 7 UKST standard fields
in the SSC region, namely in fields 382--384 and 443--446. The main
SSC core centred on A3558 at $13^h~25^m$, $-$31$^{\circ}$ is in field
444.  As each selected Abell cluster has a projected angular diameter
of 2.5 to 3.0 degrees, the wide-field UKST FLAIR-II system was an
ideal facility for this project whilst additionally permitting us to
probe the regions between the dominant clusters neglected in previous
studies. These combined fields covered an area of  182~deg$^2$
which allows us to investigate the outer limits of the SSC out to
radii of more than 10~deg (35\Mpc) from A3558.

\subsection{Input Galaxy Samples}

All target galaxies were originally obtained from the red ESO/SRC sky
survey plates of the above fields scanned by the Paris Observatory
MAMA plate-measuring machine (Guibert \& Moreau 1991; Moreau
1992). The images were classified using standard star/galaxy
separation techniques (e.g. Dickey et al. 1987, Heydon-Dumbleton et
al. 1989).  We defined a galaxy sample to a photometric limit of
$R\leq16$, corresponding (assuming a mean $B-R=1.5$) to $B<17.5$, the
nominal galaxy limiting magnitude of the FLAIR-II system (Parker \&
Watson 1995). This corresponds to an absolute magnitude of $M_B=-19$
at the Shapley distance of 200\Mpc.  This selection gave total galaxy
samples of 600--1000 per field.  All previously catalogued matches to
literature redshifts were then removed from the target samples prior
to observation.

\subsection{FLAIR-II observations}

Table~1 gives the journal of FLAIR-II observations reported here. An
interim magnetic-button fibre-ferrule system, implemented with the
71-fibre FLAIR-II plateholder, was commissioned during these
observations. This was as a proof of concept for the
recently-commissioned automated robotic fibre positioner that will
replace FLAIR-II at the UKST and known as 6dF - for 6-degree field
(e.g. Watson et al. 2000).  Note that typically 6~fibres were devoted
to the blank sky regions to facilitate sky-background subtraction. An
overall redshift success rate of 75\% was achieved from the FLAIR-II
data.

\begin{table}
\caption{Summary of FLAIR-II observations}
\label{tab_obs}
\vspace{0.2cm}
\begin{tabular}{lrrrrrrr}
\hline
Date    & Field  &  RA (J2000) Dec & time & seeing  & weather & N$_g$ & N$_r$ \\
\hline
1998 May  22 &F383B & 13:15:00 -35:15:00 &5400  & 1   &clear &58 & 43\\
1999 June 10 &F445A & 13:51:00 -30:15:00 &11000 & 2-3 &clear &65 & 53\\
1999 June 11 &F445B & 13:51:00 -30:15:00 &12000 & 3-4 &cloudy&65 & 54\\
1999 June 12 &F445C & 13:51:00 -30:15:00 &12000 & 3   &clear &65 & 52\\
1999 June 13 &F445D & 13:51:00 -30:15:00 &12000 & 2-3 &clear &65 & 41\\
1999 June 14 &F443A & 13:05:00 -30:15:00 &12000 & 1-2 &clear &73 & 21\\
1999 June 16 &F443B & 13:05:00 -30:16:00 &12000 & 1-2 &clear &73 & 32\\
1999 June 17 &F443C & 13:05:00 -30:16:00 &12000 & 2+  &cloudy&65 & 42\\
2000 June  2 &F446  & 14:14:00 -30:14:00 &15000 & 2-3 &clear &61 & 28\\
2000 June  3 &F384  & 14:03:00 -30:14:00 &15000 & 1-2 &clear &80 & 64\\
\hline
\end{tabular}
\vspace{0.2cm}

{\small Notes: N$_g$ is the number of target galaxies observed while
N$_r$ is the number of galaxies actually yielding a redshift in a
particular field configuration, not counting the spectra dominated by
Galactic stars.
}
\end{table}

The data were reduced as in Drinkwater et al.\ (1996) using the
dofibers package in IRAF (Tody 1993). Redshifts were measured for
galaxy spectra using the cross-correlation task XCSAO in RVSAO (Kurtz
\& Mink 1998) using a mixture of a dozen stellar and galaxy
templates. The galaxy templates included a number of emission line
examples.  The IRAF EMSAO utility was also used to manually check
emission line redshifts where necessary.  For spectra showing both
absorption and emission features the result (either XCSAO/EMSAO) with
the lower error was used. Note that 61 spectra were contaminated by a
dominant foreground star although we tried to remove all galaxies with
a dominant star on the line-of-sight during the fibering procedure. No
redshift was measured for these galaxies.

The complete set of FLAIR-II measurements from both papers was
carefully checked for consistency, in particular by matching the
galaxy positions to the new SuperCOSMOS catalogue (Hambly et al.\
2001a, 2001b, 2001c). During this process several errors in the data
from D99 were identified: 5 galaxies had incorrect redshifts reported
and 12 had incorrect positions, all but two of which we have been able
to correct. The 17 galaxies from D99 with incorrect measurements are
listed in Table~\ref{tab-errata}.

\begin{table}
\caption{Galaxies with incorrect measurements in D99}
\label{tab-errata}
\begin{center}
\begin{tabular}{ccrl}
\hline
Original Position$^1$ &New Position$^2$  & $cz {\pm {\Delta}cz}$ &
 Correction \\
RA (B1950) Dec &RA (J2000) Dec& (\kms)\\
\hline
 13:04:08.4 -37:06:37 & 13:06:56.1 -37:22:09 &-- & position  \\
 13:12:05.7 -33:29:36 & 13:14:52.8 -32:45:27 &-- & position  \\
 13:14:37.9 -33:39:08 & 13:17:26.2 -33:55:36 &-- & position  \\
 13:15:21.3 -31:45:59 & 13:18:08.7 -32:01:06 &-- & position  \\
 13:18:06.6 -27:47:20 & 13:20:52.8 -28:13:06 &-- & position  \\
 13:19:39.1 -33:06:29 & 13:22:28.0 -33:29:03 &-- & position  \\
 13:24:21.4 -37:00:34 & 13:27:13.1 -36:25:34 &-- & position  \\
 13:24:28.2 -36:49:27 & 13:27:10.4 -37:05:00 &-- & position  \\
 13:29:40.1 -33:54:18 & 13:32:31.3 -33:09:42 &-- & position  \\
 13:31:39.7 -36:33:12 & 13:34:33.1 -36:48:32 &   2398  17& velocity \\
 13:33:21.5 -36:38:27 & 13:36:15.3 -36:53:45 &   3897  79& velocity \\
 13:35:47.0 -34:17:07 & -- &-- & removed$^3$   \\
 13:37:15.1 -34:05:41 & 13:30:06.0 -34:21:10 &-- & position  \\
 13:40:52.6 -34:49:27 & 13:43:46.5 -35:04:30 &  14681  71& velocity \\
 13:42:33.6 -36:31:58 & 13:45:29.1 -36:46:59 &   7633  80& velocity \\
 13:43:08.2 -33:29:39 & 13:46:01.4 -33:44:39 &  24838  61& velocity \\
 13:43:32.6 -35:50:50 & -- &-- & removed$^3$   \\

\hline
\end{tabular}
\vspace{0.2cm}

{\small Notes. 1) the position given in D99; 2) the J2000
  position in this paper, corrected in some cases; 3) in two cases the
  original position was wrong, but the correct identification could not
  be determined.}
\end{center}
\end{table}

\begin{figure*}
\psfig{figure=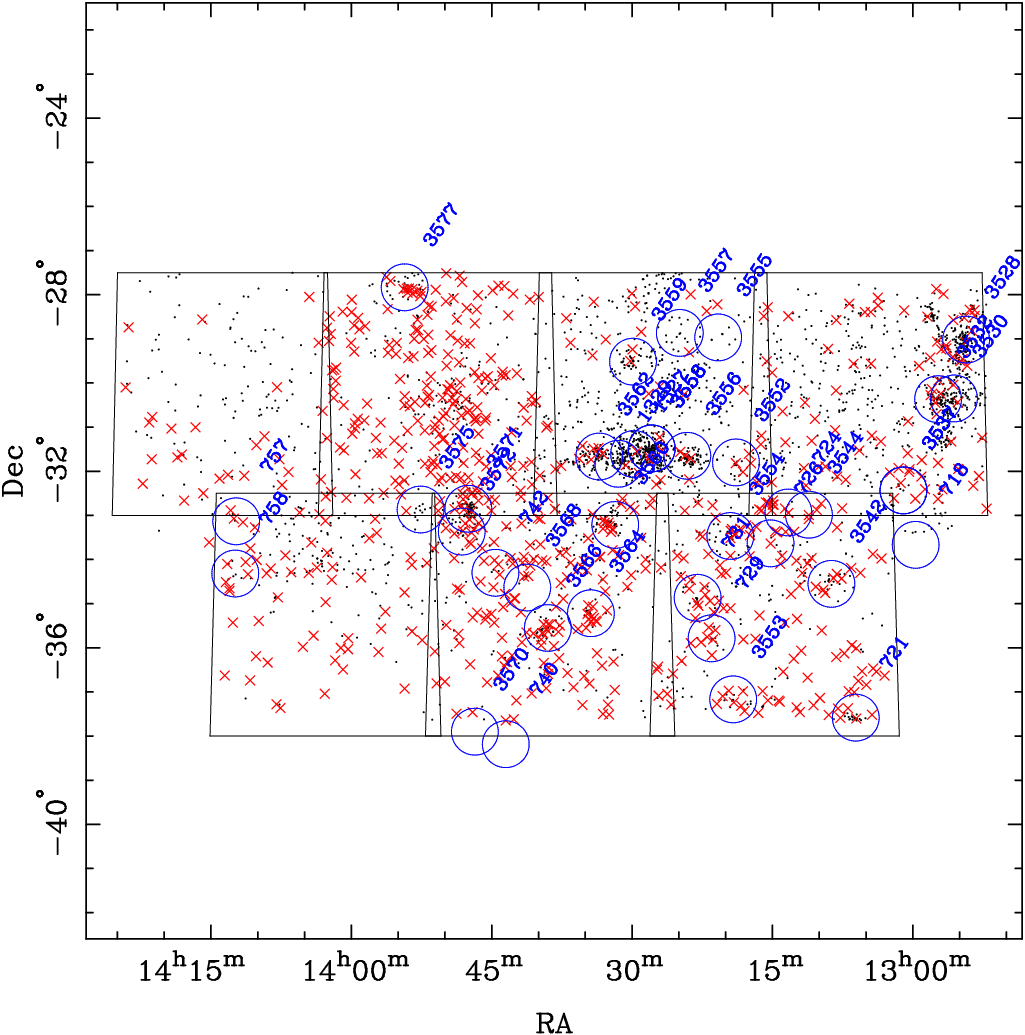,width=15cm}
\caption[]{Sky distribution of galaxies with measured redshifts in the
  Shapley Region. The galaxies observed with FLAIR-II for this work
  are plotted as crosses and those previously published as dots. Known
  galaxy clusters with redshifts overlapping the Shapley range
  ($7500<cz<18500$\kms) are indicated by large labeled circles, and
  the 5 UKST fields by large overlapping squares. Note the
  concentration of previous measurements in the clusters.}
\label{fig-sky}
\end{figure*}

Table~\ref{tab-cat} lists our complete catalogue of 710 galaxies
measured in the work reported here and our previous FLAIR-II
observations from D99. The column descriptors are given at the
end of the table. In Fig.~\ref{fig-sky} we show the projected
distribution of the observed galaxies on the sky.  We also show
galaxies with previously published redshifts and the known Abell
clusters.

\subsection{Observed Galaxy Sample and Completeness}

\begin{figure}
\hbox{
\psfig{figure=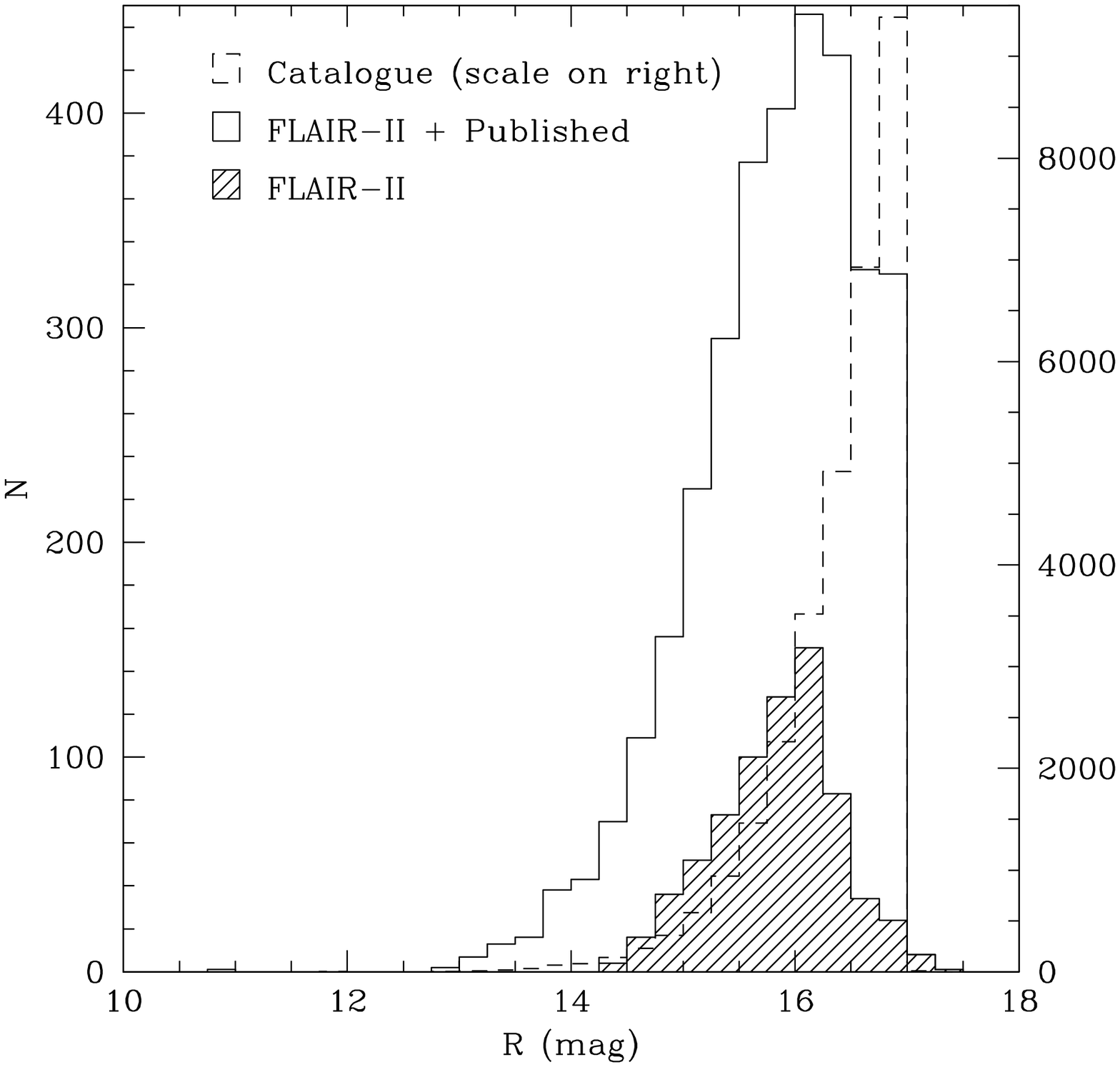,width=8cm}\psfig{figure=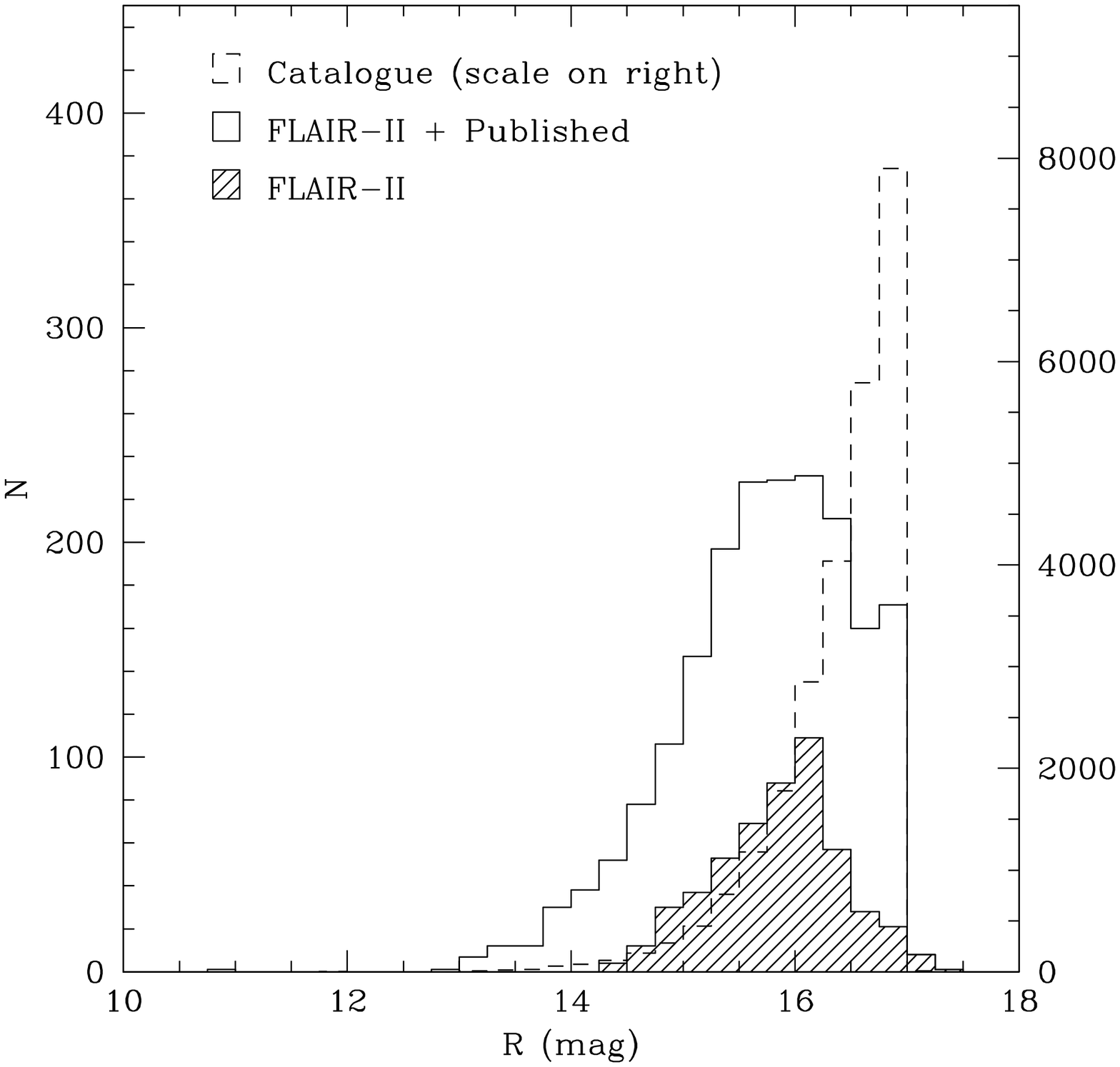,width=8cm}}
\caption[]{Magnitude distributions of observed and catalogued galaxies
in the Shapley region. Left: full samples; Right: inter-cluster region
only. Note that in each case the catalogue histogram is on a
different scale, shown on the right.}
\label{fig-mag1}
\end{figure}

For the purpose of our analysis below we wished to compare our sample
of observed galaxies to the total magnitude-limited galaxy
distribution in our survey region. We chose to use the new SuperCOSMOS
sky surveys (Hambly et al.\ 2001a,b,c) to construct the parent galaxy
catalogue for the region. This choice was motivated by the
availabililty of full image data online from the SuperCOSMOS survey as
well as some evidence that the MAMA photometry values used for the
final two fields (384 and 446) was different from the others. We
therefore quote and analyse the SuperCOSMOS R (the ``$R_1$'' value
from the U.K.\ Schmidt Telescope survey plates) magnitudes of all
galaxies in this paper. In this system the sample does not have a
sharp magnitude limit, as shown in Fig.~\ref{fig-mag1}. We compared
the magnitude distributions of the objects measured in the seven
fields: two-sample K-S tests showed that they were all consistent with
the same distribution (mean $R \approx 15.8$) except for the two final
fields---384 and 446---which were significantly fainter (mean
$R\approx 16.4$).

%

We next used the SuperCOSMOS survey to define a parent galaxy sample
over the whole region covered by our seven fields observed. Allowing
for the boundaries of the survey fields and a circular region 1 degree
in diameter we excluded around the bright star HD 123139 (at 14 06
41.0 -36 22 12, J2000), the total survey region has an area of 182
square degrees.  As in D99 we also defined a restricted
inter-cluster region by excluding regions 1 degree in diameter around
any rich galaxy clusters in the Shapley velocity range, having an area
of 159 square degrees. The excised cluster regions are shown in
Fig.~\ref{fig-sky}.

We used the parent galaxy sample to determine the completeness of our
survey as a function of limiting magnitude. This is shown in
Table~\ref{tab-samples} for both the full region and the inter-cluster
region. As in D99 we have also compiled a larger list of galaxy
measurements in the survey region based on published data (NED
searched 2003 Aug 5) and other new observations by our group (Quintana
et al.\ in preparation). The table shows that the completeness is
highest for the brighter magnitude limits, peaking at 28 per cent for
the whole sample at $m_R<16$.


\begin{table}
\caption{Galaxies observed}
\label{tab-samples}
\begin{center}
\begin{tabular}{lrrrl}
\hline
Field  & R Mag limit &  Catalogue  &  FLAIR-II  & FLAIR-II + Published \\
\hline
Full          & 16.0 &  6193 & 409 (7\%)& 1754 (28\%)\\
Full          & 16.5 & 14635 & 643 (4\%)& 2627 (18\%) [10\%]\\
Full          & 17.0 & 30949 & 701 (2\%)& 3279 (11\%) [4.0\%]\\
\hline
Full RA$<$14:50     & 16.0 &  5309 & 397 (7\%)& 1630 (31\%)\\
Full RA$<$14:50     & 16.5 & 12256 & 610 (5\%)& 2458 (20\%) \\
Full RA$<$14:50     & 17.0 & 25484 & 658 (3\%)& 3087 (12\%)\\
\hline
Inter-cluster & 16.0 &  4930 & 293 (6\%)& 1138 (23\%) \\
Inter-cluster & 16.5 & 11819 & 459 (4\%)& 1580 (13\%) [6.4\%]\\
Inter-cluster & 17.0 & 25511 & 508 (2\%)& 1911  (7\%) [2.4\%]\\
\hline
\end{tabular}
\end{center}
Notes: the completeness of each sample compared to the catalogue is
given in parentheses (integral) and brackets [differential].
\end{table}

\section{Discussion}

We have obtained a total of new 710 galaxy radial velocities from all
of our FLAIR-II observing runs. This represents a substantial
improvement in the available velocity catalogue in the SSC
particularly for the previously neglected inter-cluster regions.
In this section we use our sample to analyse the extent of the SSC and
the galaxy overdensity it represents.

\subsection{Extent of the SSC}

\begin{figure*}
\vspace{-4.0cm}
\psfig{figure=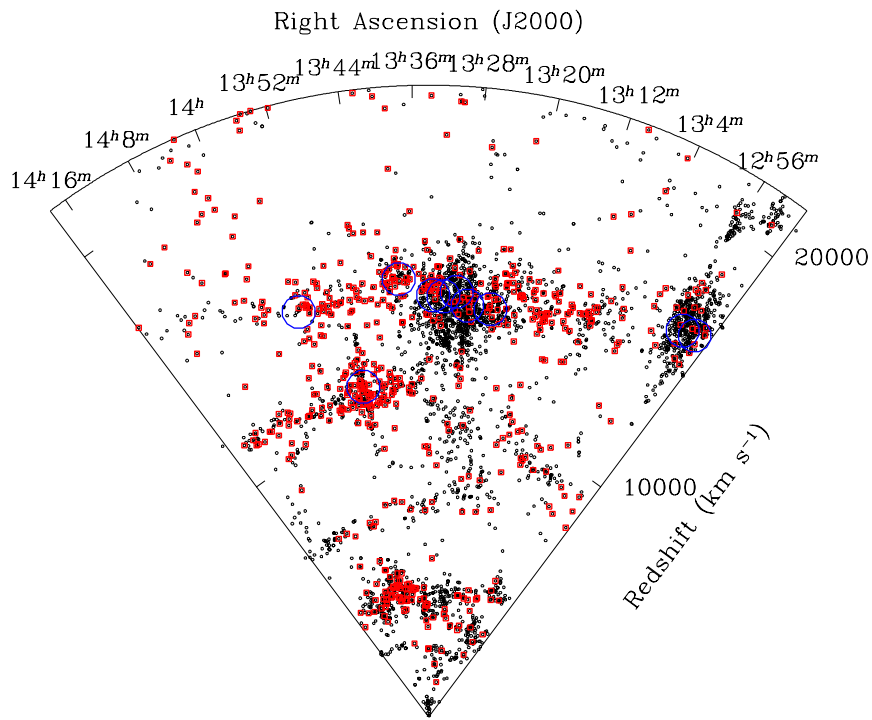,height=15cm}
\vspace{-4.0cm}
\psfig{figure=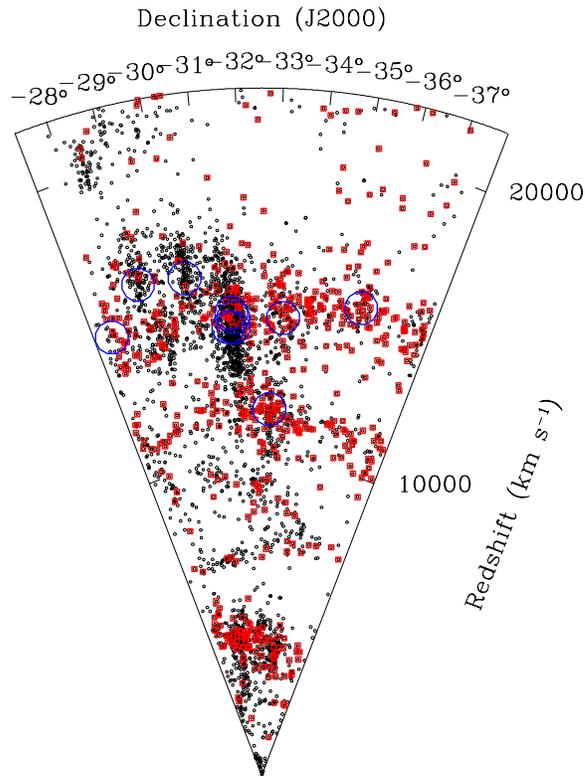,height=15cm}
\caption[]{Cone velocity diagram in right ascension from R.A.$=
12^h15^m$ to R.A.$= 14^h30^m$ (upper panel), and declination from
$-26^{\circ}$ to $-38^{\circ}$ (lower panel) for the measured galaxies
in the Shapley Supercluster. Previously published galaxies are plotted
as points; the combined new measurements from our project are
plotted as squares. Also show as large circles are the locations of
the richest galaxy clusters in this region.}
\label{fig-cone}
\end{figure*}

Fig.~\ref{fig-cone} shows the combined resulting distribution of
galaxies towards the Shapley supercluster as cone diagrams in right
ascension and declination for both available literature redshifts and
our new combined sample. The importance of the SSC in this region of
the sky is demonstrated by the fact that 65\% of galaxies in our new
combined sample belong to the SSC with velocities in the range
8000--18500~\kms.  In both plots the new data are indicated by
larger squares to emphasise their impact. It can be seen that by
probing large regions of the SSC away from the rich Abell clusters, we
have revealed significant additional structures which make complex
links with the main cluster locations.

\begin{figure}
\psfig{figure=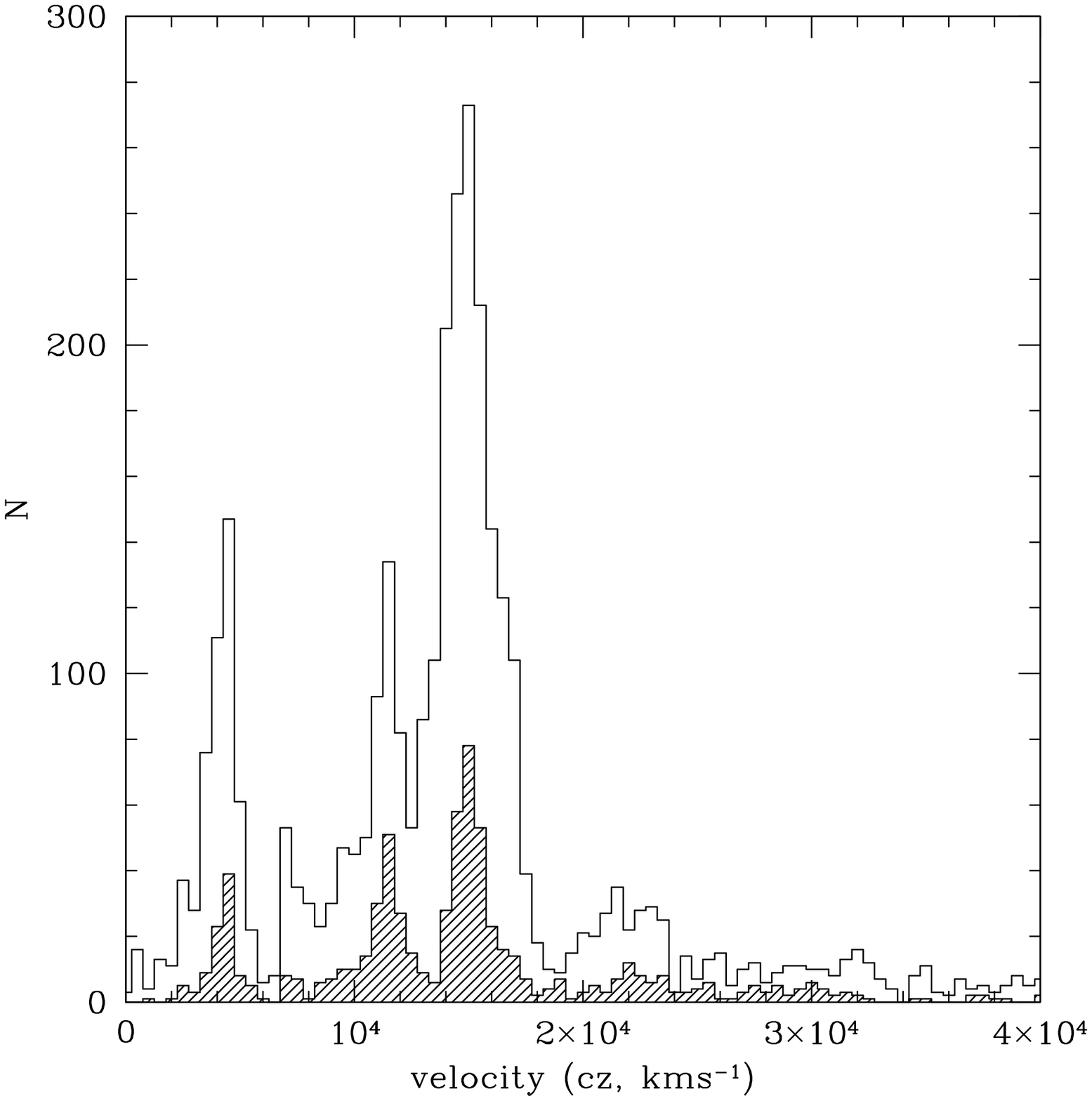,width=9cm}
\caption[]{The redshift distribution of galaxies in the SSC survey
  region. The area of sky is 182 square degrees. The shaded histogram
  shows the 710 new measurements we present here and the upper
  histogram shows those combined with all know published redshifts, a
  total of 4215 galaxies. The bin size is 500~\kms.}
\label{fig-redshift}
\end{figure}

Fig.~\ref{fig-redshift} shows the histogram of 2030 galaxy redshifts
in the direction of the Shapley supercluster with all currently
available published velocities (including our combined sample) in the
range 0~\kms $\leq$ V $\leq$ 40000~\kms with a step size of
500~\kms. The new velocities are also shown as the hatched region.


In agreement with previous works we note the presence of a prominent
foreground wall of galaxies (the Hydra-Centaurus region) at
$\overline{V}$= 4000~\kms. This distribution can be related to the
nearby cluster A3627 associated with the ``Great Attractor''
(Kraan-Korteweg et al. 1996).  Moreover, our newly measured galaxies
tend to establish a link between the Hydra-Centaurus region and the
SSC especially at the high Dec.\ end of the cone. A similar structure
was suggested by Tully et al.\ (1992).

Concerning the main SSC, our new data
revise our knowledge of its large-scale structure by measuring a large
number of galaxies away from the rich Abell clusters previously
studied.  The majority ($65\%$) of the galaxies we observed were found
to be part of the SSC, so our principal result is to show that the SSC
is bigger and more complex than previously thought. Looking at the
cone diagram in Right Ascension (Fig.~\ref{fig-cone}) we can see that
the SSC is separated into two main velocity components as previously
suggested (D99). The nearer concentration at about
$\overline{V}$=11240~\kms is located to the East of the main SSC
region at about $\overline{V}$=14850~\kms. However clumps of objects
seem to link these two main structures which were not observed
before. Care must be taken in the interpretation of the cone plots
because of the finger of God effect evident in the main SSC
concentrations due to the internal individual cluster velocity
dispersions which can lead the eye to spurious links.

A large concentration of objects at about R.A.$= 12^h54^m$ and around
$\overline{V}$= 16000-17000~\kms\  (the ``A3558'' complex; see Bardelli
et al.\ 2000) is connected to the main structure by a broad bridge of
galaxies. It can also be seen from the Declination cone diagram
(Fig.~\ref{fig-cone} lower panel) that the Southern part of the SSC
consists of two large sheets of galaxies, of which the previously
measured Abell Clusters represent the peaks of maximum density. The
more distant sheet, in particular, at $\overline{V}$= 15000~\kms is
present right across the observed region from $-27^{\circ}$ to
$-38^{\circ}$ so the true extent of this wall is currently
unknown. The Southern part of this wall may be an extension of the
plane of galaxies defined by Bardelli et al.\ (2000), although it has
the same offset of $-5$\hMpc as our earlier D99 sample when analysed
by Bardelli et al.\ in their Figure~4.

These new observations mean that we must re-appraise the conclusions
of previous papers about the overall shape of the SSC. Based on the
previously available velocity information on the main clusters it was
thought that the SSC was highly elongated and either inclined towards
us or rotating. We now reveal that the SSC extends as far as our
measurements do to the South via a broad inter-connected wall so we
find it is not elongated or flattened.  The actual situation so far
revealed is more complex, being composed of the known Abell clusters
embedded in at least two sheets of galaxies of much larger extent.

\subsection{Galaxy Overdensity}

%
%
%
%

In D99 we estimated the galaxy overdensity in an inter-cluster
region of area 44 deg$^2$ at the centre of the SSC region. Our new
observations, combined with data from the literature, allow us to
calculate the overdensity over a much larger region. We shall consider
both our full survey region of 182 deg$^2$ and our new inter-cluster
region of 151 deg$^2$.  As we are most interested in galaxies at the
distance of the SSC, the overdensity is best seen as the peaks in the
redshift histograms. 

\begin{figure*}
\hbox{\psfig{figure=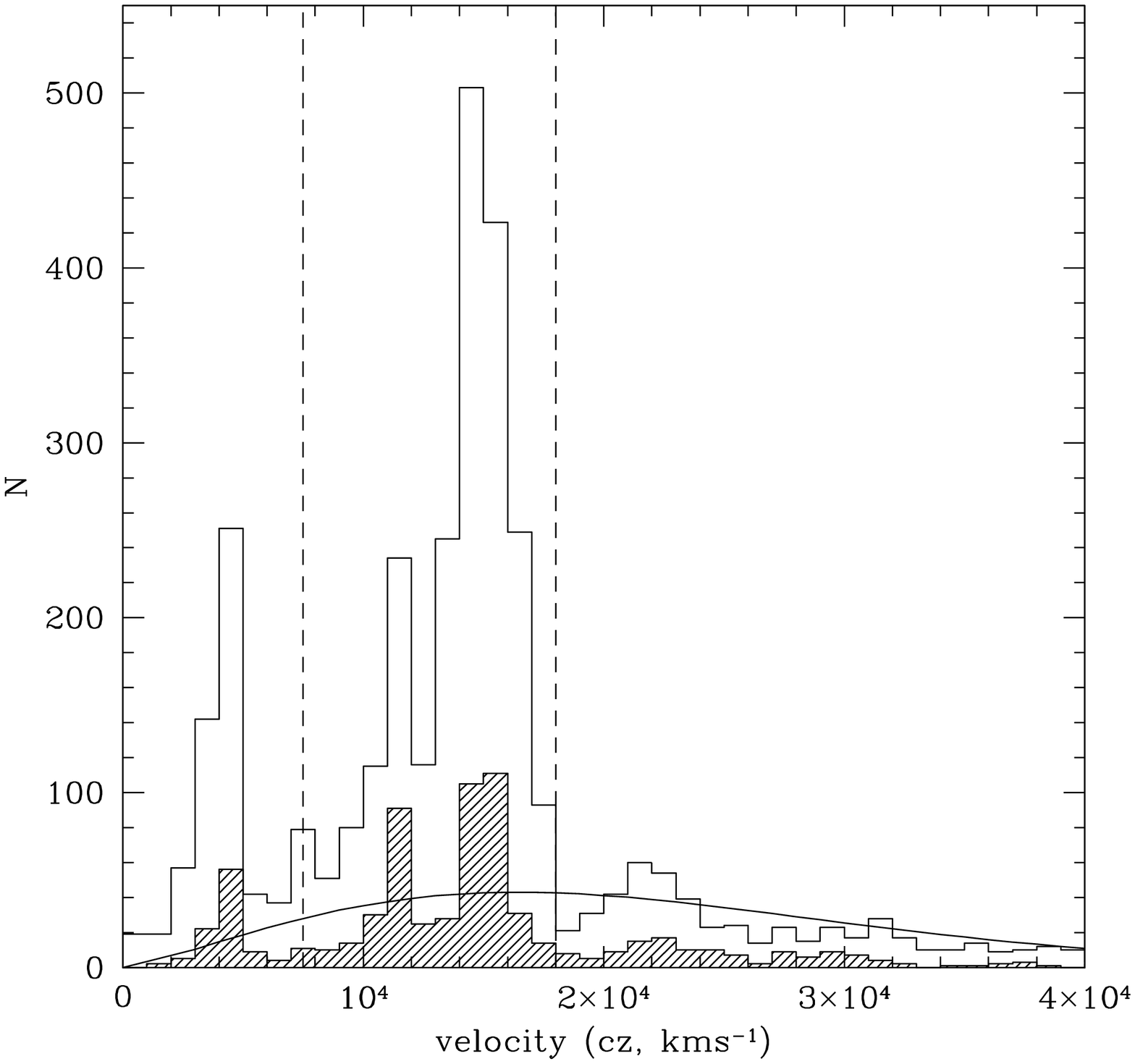,width=7cm}
\psfig{figure=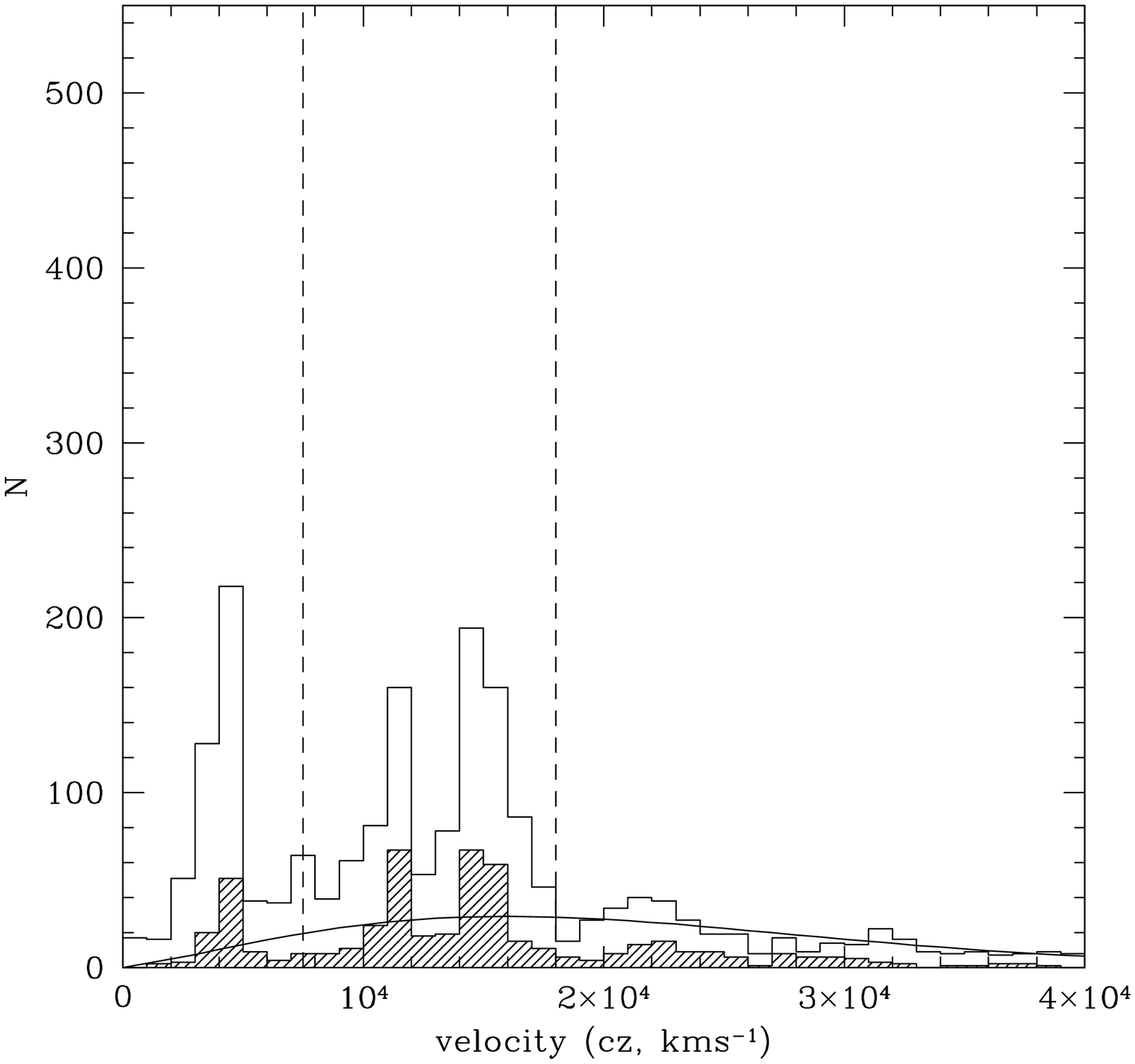,width=7cm}}
\caption[]{Redshift distributions of galaxies in the SSC region
  compared to that expected for a smooth, homogeneous galaxy
  distribution. The samples are drawn from (A) our full survey region
  and (B) the inter-cluster region. In each case the samples are
  limited to $m_R<17$. The expected distributions (Metcalfe et al.\
  1991) are scaled to the bin size, area and completeness of our
  samples. The dashed vertical lines indicate the nominal redshift
  range of the SSC, 7500--18000\kms.}
\label{fig-histo2}
\end{figure*}

In Fig.~\ref{fig-histo2} we show the redshift distributions for both
the full survey and inter-cluster regions. Also shown on both
histograms are the expected distributions for a smooth, homogeneous
galaxy distribution based on the number count data of Metcalfe et al.\
(1991). These were calculated allowing for the differential
incompleteness of each sample as a function of $m_R$ as listed in the
final column of Table~\ref{tab-samples}.

We calculated the galaxy overdensity as the ratio of the number of
observed galaxies within the nominal velocity limits of the SSC
complex (8000--18500\kms) to the number expected from the Metcalfe
counts within the same velocity limits, all with the magnitude limit
of $m_R<17$. In the full region the overdensity was $5.0\pm0.1$ and in
the restricted inter-cluster region is was $3.3\pm0.1$. This can be
compared with our result in D99 where we found $2.0\pm0.2$ over an
area of 44 deg$^{2}$.  Our overdensity values are lower than those
found by Bardelli et al.\ (2000) in a study of the central part of the
SSC. They report an overdensity of $3.9\pm0.4$ for their inter-cluster
sample and $11.3\pm0.4$ for their total sample, on scales of
$10h^{-1}$Mpc. Our data probe the distribution on much larger scales
of around 10--15 degrees corresponding to $26-40h^{-1}$Mpc.  We have
confirmed that earlier detection of a very significant galaxy
overdensity in the inter-cluster space of the SSC region. In terms of
galaxy numbers, the inter-cluster galaxies make up 46 per cent of the
2144 galaxies in the SSC velocity range, so assuming a similar mass
function they contribute a similar amount of mass to the cluster
galaxies which were the focus of most previous work on the SSC.

\section{Conclusion}

Our new observations of galaxies towards the Shapley supercluster
have, by surveying a large contiguous region including areas away from
known clusters, revealed substantial new, large, inter-connected
structures. The original SSC region has now been shown to form part of
a much larger structure extending uniformly in at least two sheets
over the whole surveyed region principally to the South of the SSC
core. We have detected an additional 426 members of the SSC in our
combined (D99 and this paper) survey area, increasing very
significantly the number of known SSC galaxies. Our new data suggest
that the SSC is at least 50\% more massive than previously thought
with a significant part of the extra mass in the closer
sub-region. The SSC must therefore has a more important effect on the
Local Group that originally envisaged. We defer a detailed calculation
of the size of this effect and a fuller analysis and interpretation of
these new data, to a subsequent paper (Proust et al, in preparation)
where some additional data will also be included.

We now plan to investigate regions well away from the SSC center
between R.A. 12h30 to 14h30 and from $-23^{\circ}$ to $-42^{\circ}$ in
order to: i) define the complete topology of the SSC, ii) analyze the
individual Abell clusters contained in the SSC and iii) determine more
precisely its gravitational effect on the Local Group. We have already
obtained the necessary extended catalogue of galaxies for this larger
region from additional red ESO/SRC SSC plates scanned by the MAMA
machine. The new 6dF replacement for FLAIR-II at
the UKST with its automated robotic fibre positioner and 150~fibres in
a $6^{o}$ field will be an ideal facility to continue our study of the
Shapley Supercluster.

\section*{Acknowledgments}
We wish to thank the UKST and AAO staff, especially Paul Cass, Malcolm
Hartley and Ken Russell. DP acknowledges receipt of a France-Australie
PICS in support of visits to Siding-Spring Observatory. We are
particularly grateful to Mike Read (ROE) for guidance with the
SuperCOSMOS survey data. We also thank Maureen Younger (UQ) for help
with the galaxy identifications.

This research was partially supported by the cooperative programme
ECOS/CONICYT C96U04 and HC thanks the FONDAP Centre for Astrophysics
and the Guggenheim Foundation for partial support.

\section*{References}


\reference Bardelli S., Zucca E., Zamorani G., Moscardini L.,
Scaramella, R., 2000, MNRAS, 312, 540

\reference Bardelli S., Zucca E., Baldi A., 2001, MNRAS, 320, 387.

\reference Dickey J.M., Keller D.T., Pennington R., Salpeter E.E., 1987, AJ, 93, 788

\reference Drinkwater M.J., Currie M.J., Young C.K., Hardy E., Yearsley J.M.: 1996,
MNRAS, 279, 595

 
\reference Drinkwater M.J., Proust D., Parker Q.A., Quintana H., Slezak E.:
1999, PASA, 16, 113 (D99)

\reference Guibert J., Moreau O., 1991, The Messenger, 64, 69

\reference Hambly N.C., et al., 2001a, MNRAS, 326, 1279

\reference Hambly N.C., Irwin M.J., MacGillivray H.T., 2001b, MNRAS, 326, 1295

\reference Hambly N.C., Davenhall A.C., Irwin M.J., MacGillivray H.T.,
2001c, MNRAS, 326, 1315

\reference Heydon-Dumbleton N.H., Collins C.A., MacGillivray H.T., 1989, MNRAS, 238, 379

\reference Kaldare R., Colless M., Raychaudhury S., Peterson B.A.,
2003, MNRAS, 339, 652

\reference Kraan-Korteweg R.C., Woudt P.A., Cayatte V., Fairall A.P.,
Balkowski C., Henning P.A.: 1996, Nature, 379, 519

\reference Kurtz M.J., Mink D.J.: 1998, PASP, 110, 943

\reference Metcalfe N., Shanks T., Fong R., Jones L.R., 1991, MNRAS, 249, 49
8

\reference Moreau O., 1992, PhD thesis, Universite Paris, 7 

\reference Parker Q.A., Watson F.G., 1995, in Wide Field
Spectroscopy and the Distant Universe, 35th Herstmonceux Conference,
ed. S.J. Maddox, \& A. Aragon-Salamanca, (Singapore: World
Scientific), 33

\reference Parker Q.A., 1997, in Wide Field Spectroscopy, 2nd conference of
the working group of IAU Commission 9 on Wide Field Imaging, 
ed. Kontizas et al., (Dordrecht: Kluwer), 25


\reference Quintana H., Ramirez A., Melnick J., Raychaudhury S.,
Slezak E.: 1995, AJ 110, 463

\reference Quintana H., Carrasco E.R., Reisenegger A., 2000, AJ, 120, 511

\reference Raychaudhury S., 1989, Nature, 342, 251

\reference Reisenegger A., Quintana H., Carrasco E.R., Maze J., 2000,
AJ, 120, 523

\reference Scaramella R., Baiesi-Pillastrini G, Chincarini G.,
Vettolani G., Zamorani G., 1989, Nature 338, 562

\reference Smoot G., et al. 1992, ApJ, 396, L1

\reference Tody D., 1993, in Astronomical Data Analysis Software and
Systems II, A.S.P. Conference Ser., Vol 52, eds. R.J. Hanisch,
R.J.V. Brissenden, \& J. Barnes, 173.


\reference Tully R.B., Scaramella R., Vettolani G., Zamorani G., 1992,
ApJ, 388, 9

\reference Watson F.G., Parker Q.A., Bogatu G., Farrell T.J., Hingley
B.E., Miziarski S., 2000, in Proc. SPIE Vol. 4008, p. 123-128, Optical
and IR Telescope Instrumentation and Detectors, Masanori Iye; Alan
F. Moorwood; Eds. (Bellingham: SPIE - The International Society for
Optical Engineering)

\bigskip

\begin{table*}[h]
\caption[]{Heliocentric redshifts for observed galaxies.}
\label{tab-cat}
\begin{flushleft}
\begin{tabular}{llllll}
\hline
RA      &   Dec    & field & $m_R$ & $v_{hel}$ & $\Delta v_{hel}$ \\
(J2000) & (J2000)  &       & (mag) & (\kms)    & (\kms)   \\
\hline
12:52:07.1&  $-$32:50:44 & 443 & 15.68&  15056 &  81  \\
12:52:35.2&  $-$31:14:36 & 443 & 16.14&  15580 &  40  \\
12:52:55.8&  $-$32:06:30 & 443 & 14.69&   8739 & 126  \\
12:53:23.8&  $-$28:20:46 & 443 & 15.81&  16403 &  60  \\
12:53:24.3&  $-$28:28:47 & 443 & 14.82&   8140 &  31  \\
12:53:33.0&  $-$32:19:44 & 443 & 15.13&   9277 &  74  \\
12:53:40.6&  $-$28:48:58 & 443 & 16.21&  16633 & 132  \\
12:53:55.1&  $-$28:20:30 & 443 & 15.91&  15938 & 166  \\
12:54:11.2&  $-$29:36:38 & 443 & 16.22&  25584 & 100  \\
12:54:23.0&  $-$29:04:17 & 443 & 16.08&  16380 &  96  \\
\end{tabular}
\end{flushleft}
\end{table*}

{\em Here we only present the first 10 lines of our 710-line catalogue
of all galaxies observed. The complete table is provided in the
electronic \LaTeX\  input file of the paper and as a separate text file.}

\end{document}